# Fast Yet Quantum Efficient Few-Layer Vertical $MoS_2$ Photodetectors

*David Maeso*[a], *Andres Castellanos-Gomez*[b], *Nicolas Agraït*[a,c,d] *and Gabino Rubio-Bollinger*[a,d]

[a]Departamento de Física de la Materia Condensada, Universidad Autónoma de Madrid, 28049 Madrid, Spain.

[b]Materials Science Factory, Instituto de Ciencia de Materiales de Madrid (ICMM), Consejo Superior de Investigaciones Científicas (CSIC), Sor Juana Inés de la Cruz 3, 28049 Madrid, Spain.

[c]Instituto Madrileño de Estudios avanzados en Nanociencia (IMDEA-Nanociencia), 28049 Madrid, Spain.

[d]Condensed Matter Physics Center (IFIMAC), Universidad Autónoma de Madrid, E-28049 Madrid, Spain.

E-mail: gabino.rubio@uam.es

*Semiconducting two-dimensional (2D) materials, such as molybdenum disulfide ($MoS_2$) and other members of the transition metal dichalcogenide (TMDC) family, have emerged as promising materials for applications in high performance nanoelectronics exhibiting excellent electrical and optical properties. Here, highly efficient photocurrent generation is reported in vertical few-layer $MoS_2$ devices contacted with semitransparent metallic electrodes. The light absorption of the device can be improved by fabricating vertical photodevices using few-layer flakes, achieving a photoresponse of up to 0.11 A/W and an external quantum efficiency of up to 30%. Because the vertical design, the distance between electrodes can be kept in the range of a few nanometers, thus substantially reducing collection time of photogenerated carriers and increasing the efficiency. The wavelength dependent photocurrent (PC), photoresponsivity ($\mathfrak{R}$) and external quantum efficiency (EQE) are measured over the photon energy range from 1.24 to 2.58 eV. Compared to previous in-plane and vertical devices, these vertical few-layer $MoS_2$ photodevices exhibit very short response time, ~60 ns and a cutoff frequency of 5.5 MHz, while keeping a high photoresponse.*

2D materials are a very interesting family of materials because of their strong light-matter interaction,[1] strong electron-hole confinement,[2] large surface-to-volume ratio, ultrathin planar structure, transparency, flexibility and extreme bendability. TMDCs are 2D materials, where the individual layers are held together by van der Waals forces, with an electronic band gap in the visible part of the spectrum making them strong candidates for next-generation photodetectors.[2-6] While most research on $MoS_2$ has been focused on in-plane monolayer optoelectronic devices[7-10] because of the benefits of a direct bandgap, their performance for optoelectronic devices can be limited by a relatively long semiconducting channel and a modest light absorption. On the other hand, vertical heterostructures formed by stacking few-layer $MoS_2$ nanosheets with other TMDCs have been studied for solar energy conversion achieving high power-conversion efficiencies.[11-13] Different $MoS_2$ device configurations have been studied in order to improve their performance as optoelectronic detectors such as graphene/$MoS_2$,[14] $MoS_2$-metal junctions[15-17] and monolayer and few-layer TMDC stacks.[18-21] In addition, in-plane $MoS_2$ optoelectronic devices require a gate bias, from -70 to 40 V,[3] and large drain-source bias voltages, up to 8 V,[3] in order to obtain high sensitivity, implying high power consumption and a possible breakdown of the photodevice.[22] Moreover, recent studies in layered perovskite photodetectors,[23, 24] monochalcogenide-dichalcogenide heterostructure photodevices,[25] synthesis of nanocrystals on top of 2D materials[26, 27] to improve the perfomance of photodetectors, wafer-scale synthesis of monolayers [28-30] as well as single atom confined two-dimensional materials as catalysts for various electrochemical applications,[31, 32] have attracted interest because of their optoelectronic properties.

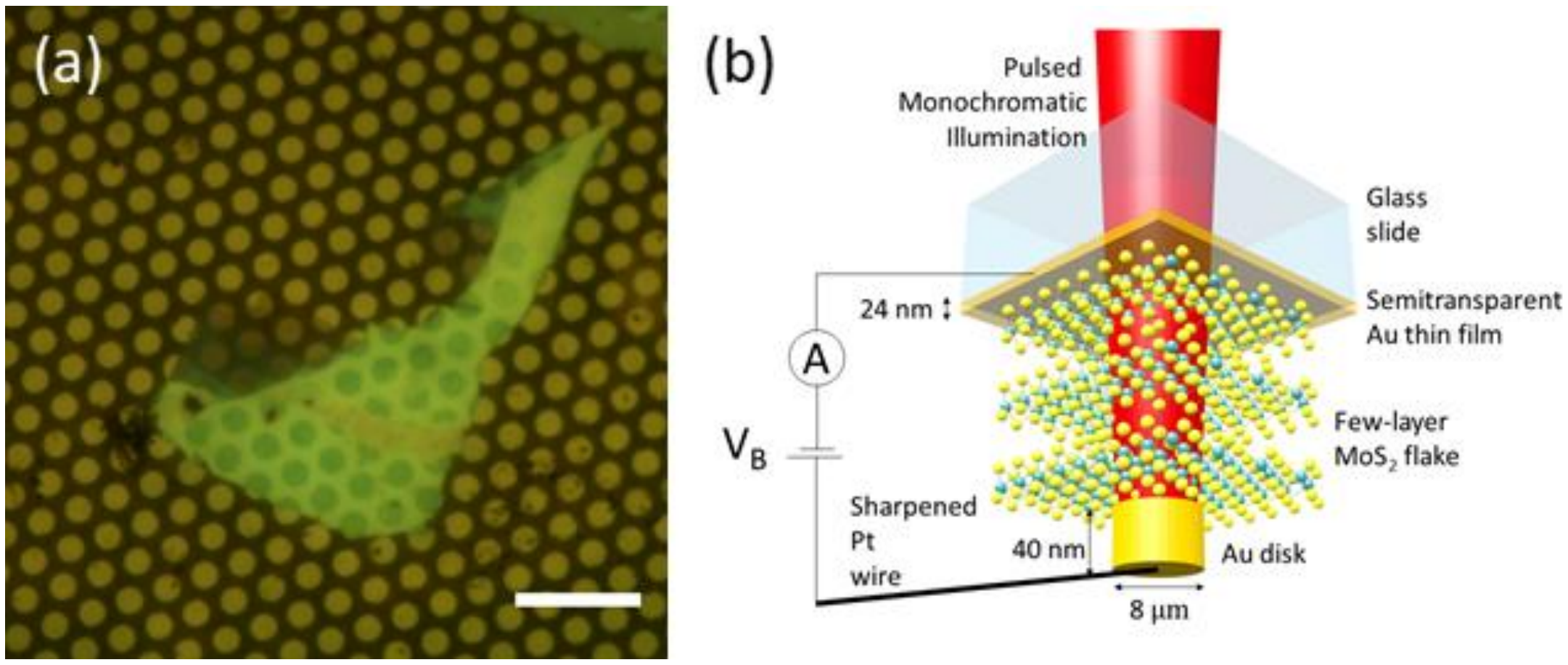

**Figure 1.** (a) Optical microscopy image of the vertical $MoS_2$ photodevices. A few-layer $MoS_2$ flake with different thicknesses is observed in the center of the image on the Au substrate. Au disk electrodes, 8 µm in diameter and 40 nm thickness, have been deposited on the $MoS_2$ flake. Scale bar is 20 µm. Note that both $MoS_2$ flake and Au disks can be distinguished through the semitransparent Au thin film. (b) Schematic of the experimental set-up for the measurement of photocurrent *I-V* characteristics. A bias voltage is applied to a single Au disk, which is in electrical contact with a sharpened Pt wire that can be precisely positioned on a selected disk. We illuminate the photodevices using a pulsed laser, which is filtered by a single grating monochromator. The monochromatic beam will be partially transmitted through the semitransparent thin film electrode and will generate photocarriers in the device.

Here, we fabricate few-layer devices that take advantage of the increased optical density of few-layer flakes and a fast collection of the photogenerated carrier pairs because of the short semiconductor channel length, which is fully encapsulated between semitransparent metallic electrodes. A semi-transparent gold electrode is fabricated depositing a 24 nm thick Au film on a glass slide substrate at a rate of 1.5 Å/s by thermal evaporation. $MoS_2$ nanosheets are exfoliated by micromechanical cleavage, under ambient conditions, using the Scotch tape method[33, 34] from bulk $MoS_2$ (Molly Hill mine, Quebec, Qc, Canada) with Nitto tape (SPV224 clear; Nitto Denko) using poly-dimethylsiloxane (PDMS) viscoelastic stamps (WF-X4 film; Gel-Pak). We transfer the nanosheets onto a gold substrate by a dry deterministic transfer method using an XYZ micromanipulator.[35] Subsequently we deposit Au disk electrodes on top of the nanosheet, with a diameter of 8 µm and 40 nm thickness, by stencil lithography using a shadow mask (G2000HAN; TED PELLA). An optical microscopy image of vertical $MoS_2$ photodevices is shown in **Figure 1**a.

We study electron transport in our devices measuring characteristic *I-V* curves. These *I-V* curves are measured under ambient conditions using a high speed low noise current amplifier at a gain of $10^5$ V/A with an acquisition card (PCIe-6363; National Instruments) at 2 MHz sampling rate. We illuminate the device using a modified optical microscope (Nikon Eclipse LV100) to focus the light with a 20X objective (NA = 0.45) on an area restricted to a single disk electrode. Our illumination source is a supercontinuum white laser (SuperK Compact; NKT) filtered through a single grating monochromator (IMS3011B; Optics Focus) to provide monochromatic light over the 480-1000 nm wavelength range, which corresponds to a photon energy range from 1.25 to 2.60 eV, with a band pass of less than 1 nm. The mean optical power incident on the device was measured at the sample using a power meter (PM100D; Thorlabs) with a calibrated silicon photodiode (S120C; Thorlabs). A schematic of the experimental set-up for the measurement of photocurrent *I-V* characteristics is shown in Figure 1b.

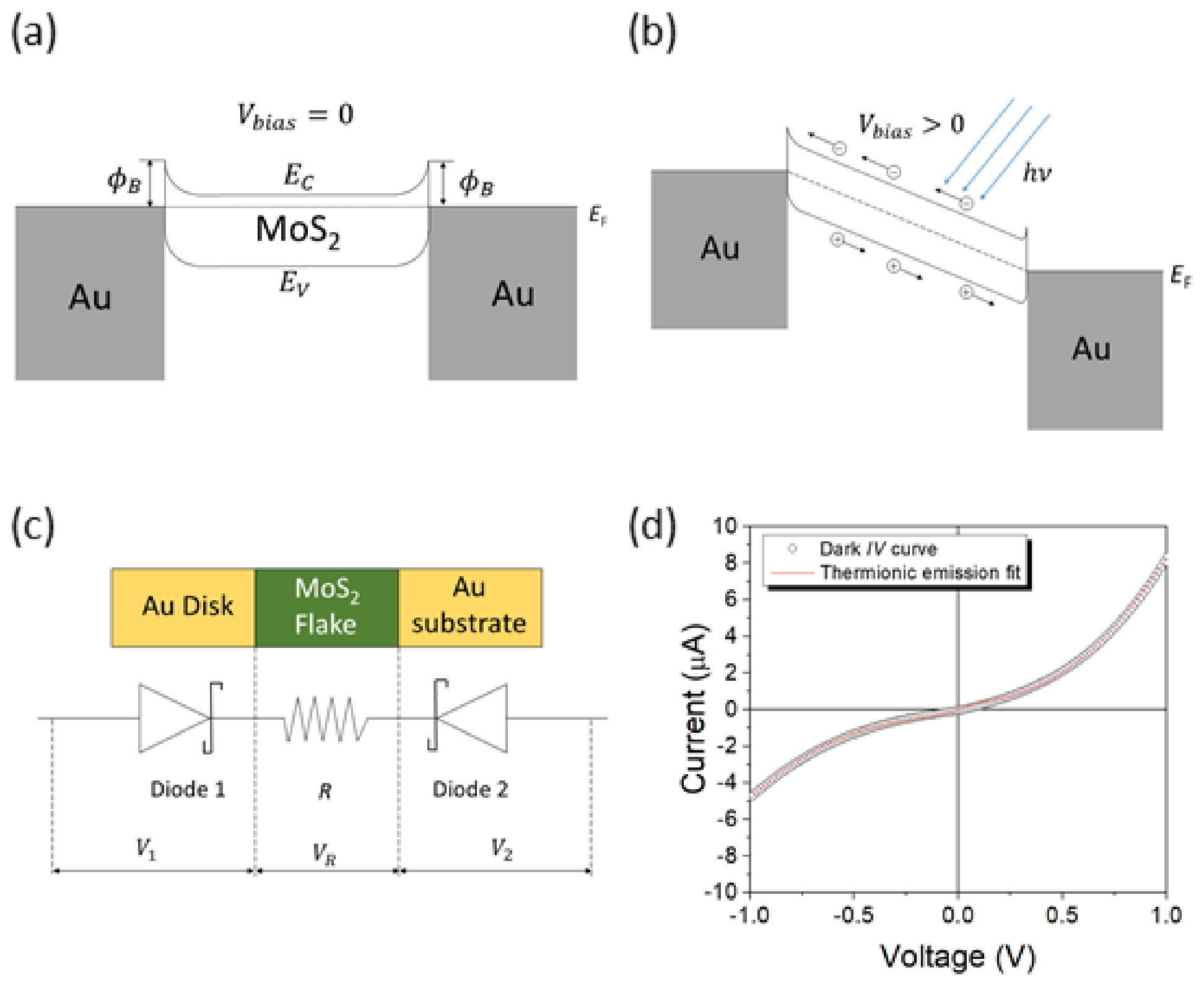

**Figure 2.** (a) A schematic band diagram of the vertical few-layer $MoS_2$ device without illumination and bias voltage. (b) A schematic band diagram of the vertical $MoS_2$ photodevice under monochromatic illumination and with applied bias voltage. (c) The schematic of the circuit where the metal-semiconductor-metal configuration can be modelled as two Schottky diodes back-to-back, one forward biased and other reverse biased, in series with a resistor. (d) An experimental dark *I-V* characteristic curve is plotted with the thermionic emission fit. A good agreement between the experimental and fitted data suggests that the electron transport is fully supported by the thermionic emission theory.

We select 20 nm thick flakes because their increased optical density over monolayer $MoS_2$ and their effective charge separation region still extends to the entire volume. As a result, an external quantum efficiency of up to 30% is achieved. Since transmittance of the semitransparent electrode is around 40-60% in the visible range (see Supporting Information for experimental data) higher values of EQE could be achieved using a lower absorbance material for the electrode, such as indium-tin-oxide (ITO), which is commonly used in solar cell technology. As the metal electrodes can be much closer in vertical devices than in in-plane devices, a shorter channel length results in a much shorter optical response time. The top and bottom electrodes protect the $MoS_2$ semiconducting channel from the ambient contaminants, reducing unintentional surface charge. Additionally, a thermal annealing in vacuum at 240 °C during 3 hours removes adsorbates from the metal-semiconductor interfaces.

When a semiconductor is brought into contact with a metal, a Schottky barrier is formed at the interface, whose height is, according to the Schottky-Mott rule,[36, 37] given by $\phi_{\mathrm{B}} = \phi_{\mathrm{metal}} - \chi_{\mathrm{semiconductor}}$, where $\phi_{\mathrm{metal}}$ is the metal work function and $\chi_{\mathrm{semiconductor}}$ is the semiconductor electron affinity. In our device, since $\phi_{\mathrm{Au}}$ = 5.1 eV [38] and $\chi_{\mathrm{MoS_2}}$ = 4.0 eV, [39] the Schottky barrier energy is 1.1 eV. A band diagram is shown in **Figure 2**a of the Au-$MoS_2$-Au system at equilibrium when no bias voltage is applied. We study electron transport in our vertical $MoS_2$ devices without illumination, measuring the *I*-*V* characteristic curve, which follows a rectifying behaviour (Figure 2d). A Metal-Semiconductor-Metal (MSM) device can be modelled as two metal-semiconductor diodes connected back-to-back in series with a resistance *R*, see Figure 2c. [40, 41] Using thermionic emission theory, [42-45] one Schottky diode is forward biased while the other one is reverse biased. When a bias voltage is applied ($V_{bias}$), voltage drops occur at the first metal-semiconductor interface ($V_1$), the series resistance ($V_R$) and the second metal-semiconductor interface ($V_2$) and one must have $V_{bias}=V_1+V_R+V_2$. The current $I_i$ through the diodes can be expressed as, including image force effects,[36] $I_i = I_{0i}\exp(qV_i / \eta_i k_B T)[1-\exp(-qV_i/ k_B T)]$ where $I_{0i} = A_i A^* T^2 \exp(-q\phi_i/ k_B T)$, $q$ is the electron charge, $\eta_i$ is the diode ideality factor, $k_B$ is the Boltzmann constant, $T$ is the absolute temperature, $A_i$ is the effective contact area, $\phi_i$ is the Schottky barrier and $A^*$ is the Richardson constant, $A^* = 4\pi q m^* k_B^2/h^3$. Here $h$ is the Planck constant and $m^*$ is the effective mass.[46] In this model, the free parameters are: Schottky barriers $\phi_1$ and $\phi_2$, series resistance $R$ and ideality factors $\eta_1$ and $\eta_2$. These unknown parameters are obtained from a fit to the measured *I*-*V* curve. The non-linear equation system $I_1=I_R$, $I_R=V_R/R$, $I_1=I_2$ is solved numerically. Taking into account that the $MoS_2$ effective mass is $m^* = 0.5 \cdot m_e$,[47] where $m_e$ is the electron mass, the effective areas are $A_1=A_2=50$ $\mu m^2$ and the absolute temperature is $T$=300 K, we solve the non-linear equation system. From the fit, we obtain the Schottky barriers $\phi_1 = 0.44$ eV and $\phi_2 = 0.44$ eV, the ideality factors $\eta_1 = 1.07$ and $\eta_2 = 1.12$ and a series resistance $R = 3.38 \cdot 10^4$ Ω. Taking into account that there is an excellent agreement between experimental and fitted *I*-*V* characteristics (see Figure 2d) and the ideality factors extracted from the fit are close to 1, the electron transport is fully explained by the thermionic emission theory. The Schottky barrier heights extracted from the fit are the same but the ideality factors are slightly different. This small asymmetry is due to the interfaces and we attribute it to the fabrication process. The deterministic transfer of $MoS_2$ to the semi-transparent electrode is performed under environmental conditions, while the disk electrodes are deposited by thermal evaporation in a high vacuum chamber.

We have obtained Schottky barriers slightly lower than those given by the Schottky-Mott rule, similar to those found in previous studies of metal-semiconductor interfaces, which have reported values from 0.12 to 0.53 eV.[48, 49] This barrier reduction is attributed to a Fermi level pinning as a result of two main effects: a metal work function adjustment due to interface dipole formation resulting from a notable interface charge redistribution and the presence of gap states at the weakened Mo-S intralayer bonding as a consequence of metal-S interaction.[48, 50] Previous studies of vertical few-layer $MoS_2$ devices with different metal electrodes found similar values (from 0.31 to 1.81 eV) for the Schottky barriers and ideality factors. [43, 44] We observe a barrier reduction from the Schottky-Mott rule, from 1.1 eV to 0.44 eV, which might be attributed to Fermi level pinning, and which can also have an impact on the found ideality factors for the interfaces. The result is that the characteristic IV curve is not an ideal Schottky junction curve, but tends to partially evolve into ohmic-like.

We illuminate the devices using a monochromatic pulsed laser source with a pulse duration of less than 2 ns. Under illumination, there is a current due to the photogenerated electron-hole pairs (Figure 2b). We measure the *I*-*V* curve and we identify sudden pulses of current correlated with the illumination pulses. Dark and illuminated *I*-*V* curves are shown in **Figure 3**a, where the illuminated *I*-*V* curve is the photocurrent at the pulse maximum. We find that at zero bias voltage there is a current flowing through the device as for photodiodes[51] and solar cells[11] at zero-bias voltage. In order to compare our data with previous reports, mean photocurrent is calculated dividing the summation of the charge photogenerated, by the time between optical pulses: Mean PC $= \sum_i^N (I_i - I_{\mathrm{dark}})\, T_S / T_{\mathrm{pulses}}$. Here $I_i$ is the current under illumination, $I_{\mathrm{dark}}$ is the current without illumination, $T_S$ is the sampling period of the data acquisition and $T_{\mathrm{pulses}}$ is the time between pulses. Pulsed PC, the maximum value of the photocurrent at each bias voltage, is plotted in Figure 3b under different bias voltages for different photon energy excitation. We find a square root like dependence of the photocurrent *vs.* voltage.[52] No photocurrent is generated when photon energy is below 1.8 eV, that is, below the $MoS_2$ gap energy, and we identify a maximum of PC at 2 eV. We also measure the mean incident power at the photodevices and calculate two figures-of-merit: mean photoresponsivity ($\Re$) and mean external quantum efficiency (EQE) as a function of illumination wavelength.[53] Responsivity characterizes the electrical output per optical input and is defined as $\Re = \text{Mean PC}/\text{ Mean } P_{\text{light}}$. Mean $P_{\text{light}}$ is the incident light power. The EQE is the ratio between the number of charges collected by the photodevice and the total number of incident photons and is given by $\text{EQE} = n_e/n_{photon}^{total} = \Re\, h\nu/q$ , where $\nu$ is the photon frequency. $\Re$ and EQE spectra (Figure 3c) exhibit the same characteristics: both $\Re$ and EQE are close to zero for photon energies below 1.7 eV and two distinct peaks related to the $MoS_2$ excitons A and B are identified, the exciton A at 1.85 eV and the exciton B at 2.05 eV. These energy dependent spectra are related to the optical absorption in few-layer $MoS_2$, which is dominated by direct transitions at the K-point of the Brillouin zone (1.7 eV) instead of transitions related to the indirect bandgap near the Γ-point (1.2 eV). [4, 6] Maximum values of both $\Re$ and EQE are found at higher energies, in the proximities of exciton C. The incident power dependence of the photoresponse is studied focusing a 640 nm LED on the device. The photon energy (1.94 eV) is close to a maximum of the photoresponse. **Figure 4**a shows a log-log plot of both PC and $\Re$ *vs*. power density. The relationship between PC and optical power *P* follows $I_{\text{photo}} \propto P^{\gamma}$,[54] where $\gamma$ ranges from 0.5 to 1 depending on the carrier recombination mechanism. $\gamma = 0.5$ implies a recombination of statistically independent and oppositely charged carriers, leading to a sublinear increase of the photocurrent at higher light intensities.[54] A linear fit of the PC log–log plot shows a slope of 0.51, suggesting that both photogenerated carriers contribute to the total PC.

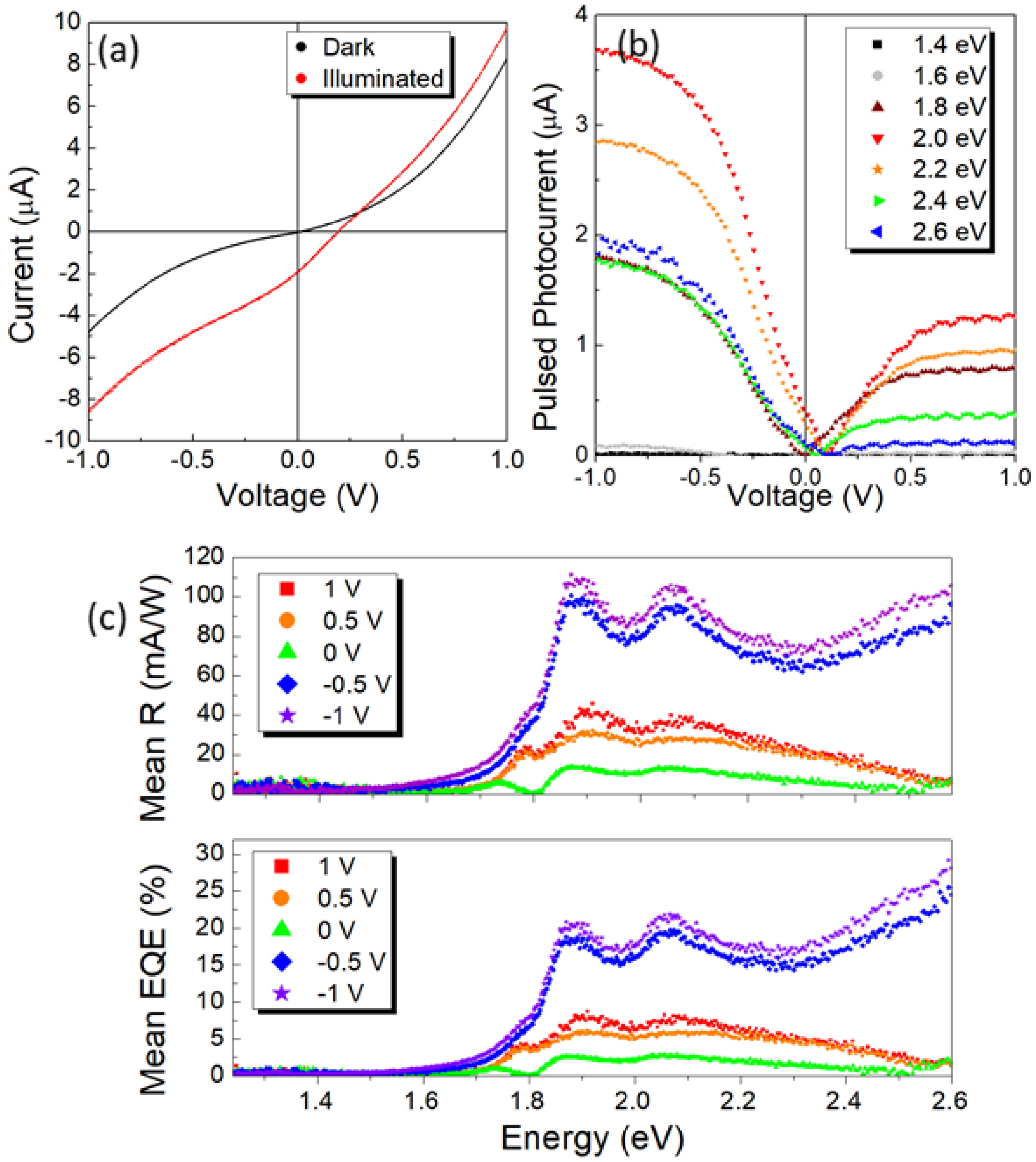

**Figure 3.** (a) Characteristic *I-V* curve of a vertical few-layer $MoS_2$ device, 20 nm thickness, without illumination (black) and under illumination (red) with 2 eV photon energy. (b) Pulsed photocurrent *vs.* bias voltage for different photon energies where a square root like dependence is observed. (c) Mean $\Re$ (top panel) and EQE (bottom panel) spectra for different bias voltages reaching a $\Re$ of up to 110 mA/W and an EQE of up to 30%. Two dominant peaks related to the A and B excitons are identified.

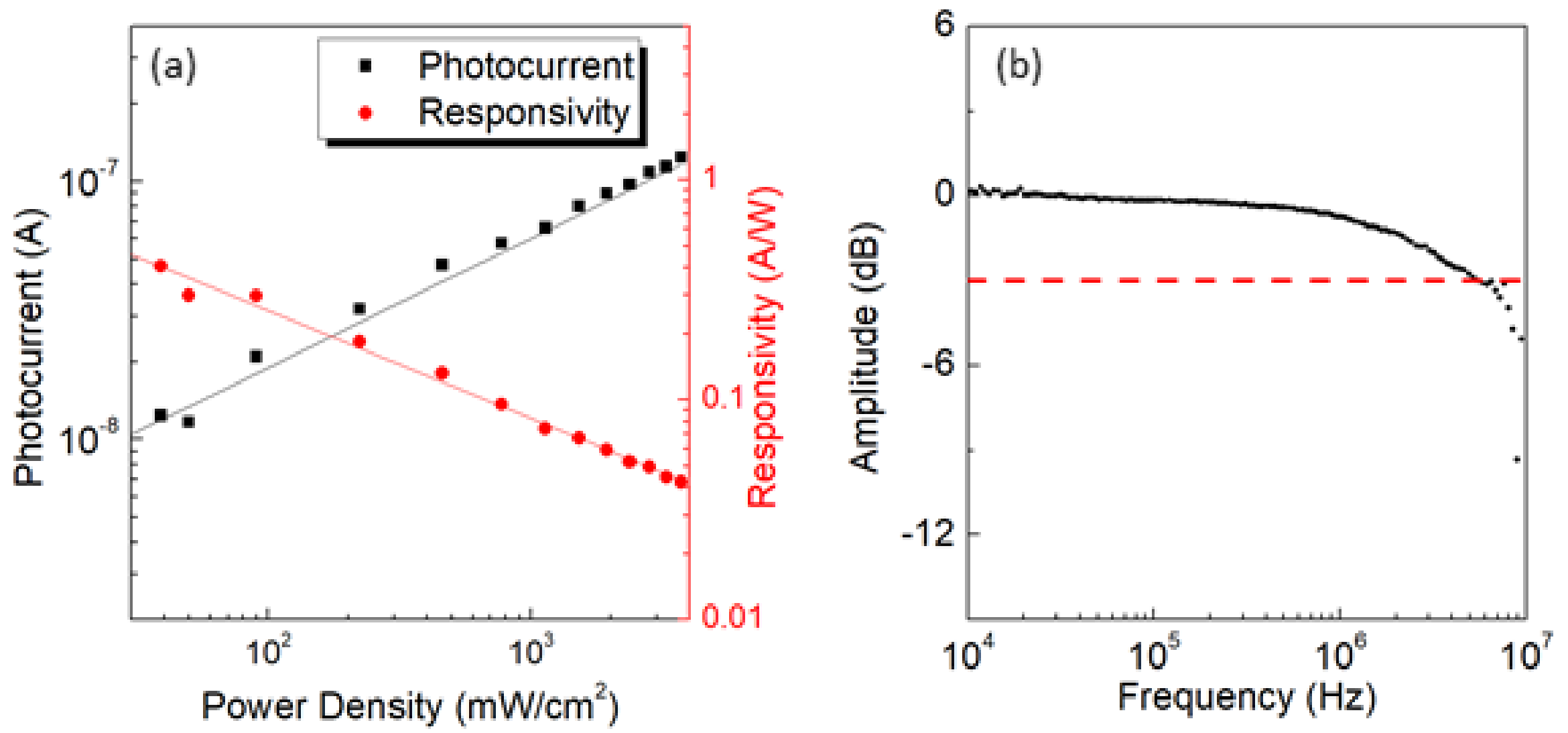

**Figure 4.** (a) Log-log plot of photocurrent and photoresponsivity *vs.* illumination power density. Photocurrent has an exponential dependence with the optical power as $I_{photo} \propto P^{\gamma}$. These exponents, $\gamma = 0.51$ for PC and $\gamma = -0.49$ for $\Re$, are calculated from the fit of the data and suggest that both photogenerated electrons and holes contribute to the total PC. (b) Normalized frequency response of the photocurrent, which shows a –3 dB cutoff frequency at 5.5 MHz and a response time ($t_r$=0.35/$f_{cutoff}$) of ~ 60 ns.

We characterize the response time of the device modulating the illumination intensity delivered by the 640 nm LED and detect the generated PC using a lock-in amplifier (HF2LI; Zurich Instruments) and high speed low noise current amplifier (DHPCA-100; FEMTO) with 14 MHz bandwidth. We measure the frequency dependent photocurrent, Figure 4b, and find a -3 dB cutoff frequency of 5.5 MHz, which corresponds to a rise time of ~ 60 ns ($t_{\text{r}} = 0.35/f_{\text{cutoff}}$). Our devices reach this high speed due to their very short semiconducting channel, favouring a fast collection of the majority of photogenerated carriers. The response time is significantly shorter than in other vertical devices previously reported (7 μs). [51]

Our vertical optoelectronic detectors exhibit a photoresponsivity of 0.11 A/W, an EQE of up to 30%, a response time of ~60 ns and a bandwidth of 5.5 MHz. In few-layer photodevices [55-57] and in-plane monolayer,[3] high responsivities of up to $10^5$-$10^8$ A/W have been reported but with much slower response time, ranging from 10 ms to 10 s, due to the photogating effect.[3, 58] Other previous studies on in-plane $MoS_2$ photodetectors have reported responsivity (0.57 to 1 A/W) comparable to that of our device, but at the expense of a response time three orders of magnitude longer (40 to 70 μs).[59, 60] See Supp. Information for a detailed comparison of $MoS_2$-based photodevices. Our photodetectors achieve a balance between responsivity and response time, reaching a substantially high responsivity (larger than 0.1 A/W) while attaining a fast response time of ~60 ns.

In summary, we fabricate vertical Au-$MoS_2$-Au photodevices that are fast and highly efficient photodetectors. The metal-semiconductor interfaces form thermionic emission Schottky barriers. In such a vertically stacked structure, one can tailor the semiconducting channel length to a desired optical density while keeping the electrode separation in the nanometer range. The entire volume of the few-layer $MoS_2$ contributes to generate photocarriers, which are efficiently collected before recombination due to the reduced length of the semiconducting channel. These advantages make our photodetector reach much faster photoresponse than in-plane devices. Using pulsed monochromatic illumination, we study the wavelength dependent mean photocurrent, photoresponsivity and external quantum efficiency spectra. Above a photon energy of 1.7 eV the photodetector shows a large EQE of up to 30%, a $\mathfrak{R}$ of 0.11 A/W, a response time of 60 ns and a bandwidth of 5.5 MHz, making these devices fast and highly sensitive detectors in the visible light range.

**Acknowledgements**

DM, NA and GR-B acknowledges funding from MINECO through grants MAT2017-88693-R, MAT2014-57915-R, BES-2015-071316 and the María de Maeztu Programme for Units of Excellence in R&D MDM-2014-0377.

# Supporting Information

## Fast yet quantum efficient few-layer vertical $MoS_2$ photodetectors

*David Maeso*[a], *Andres Castellanos-Gomez*[b], *Nicolas Agraït*[a,c,d] *and Gabino Rubio-Bollinger*[a,d*]

[a]Departamento de Física de la Materia Condensada, Universidad Autónoma de Madrid, 28049 Madrid, Spain.
[b]Materials Science Factory, Instituto de Ciencia de Materiales de Madrid (ICMM), Consejo Superior de Investigaciones Científicas (CSIC), Sor Juana Inés de la Cruz 3, 28049 Madrid, Spain.
[c]Instituto Madrileño de Estudios avanzados en Nanociencia (IMDEA-Nanociencia), 28049 Madrid, Spain.
[d]Condensed Matter Physics Center (IFIMAC), Universidad Autónoma de Madrid, E-28049 Madrid, Spain.

E-mail: gabino.rubio@uam.es

Contents:

## 1. Semitransparent electrode transmittance spectra

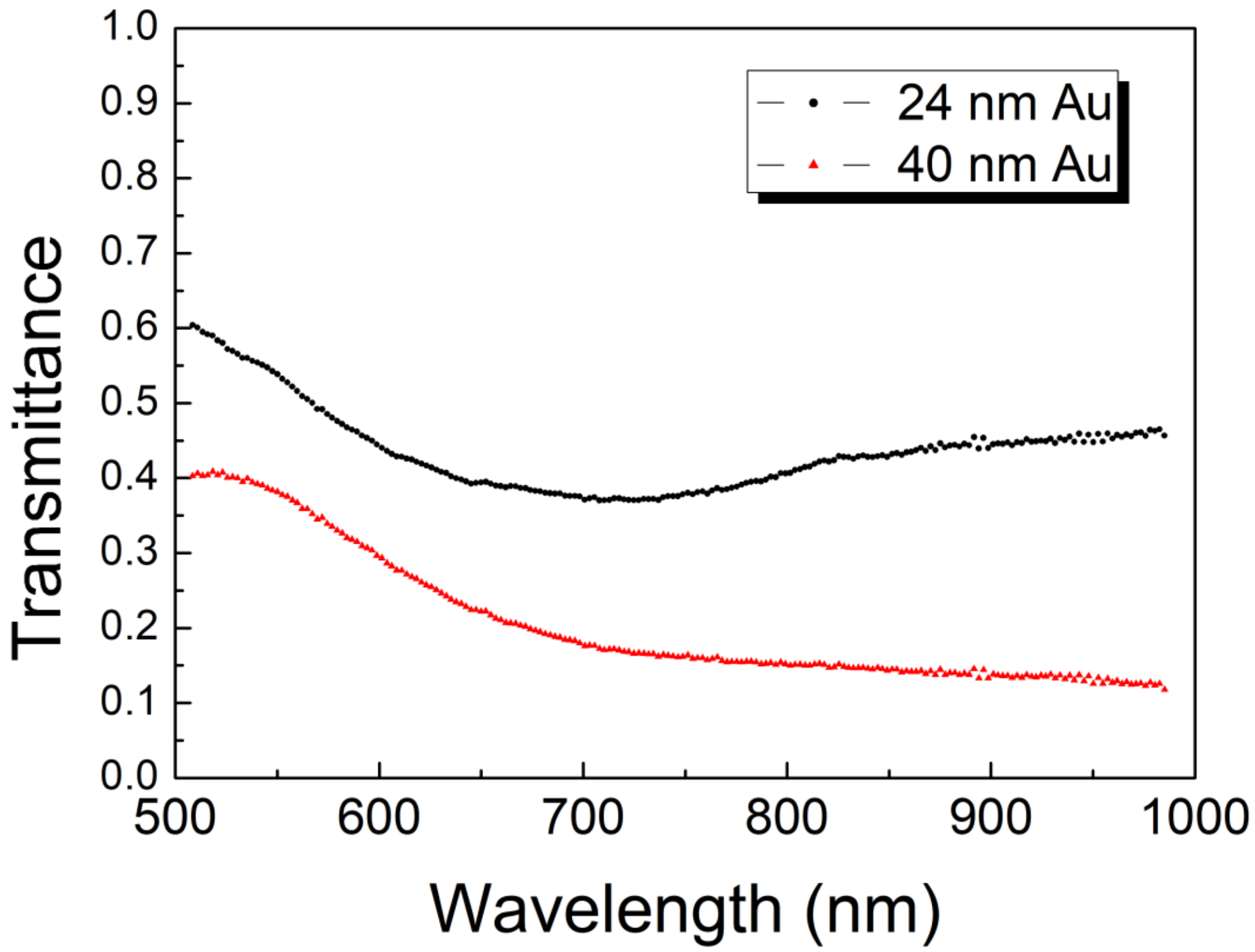

**Figure S1**. Transmittance spectra of semitransparent gold thin films used as substrate (black dots) and disks (red triangles).

We measure micro-transmittance[1] in both semitransparent electrodes using a 2.5 µm fiber pinhole and illuminating with an infrared enhanced white light (OSL2 and OSL2BIR, Thorlabs) in transmission mode. The transmitted light is collected by the same objective that we have used to illuminate the photodevices (Nikon 20X, 0.45 NA) and we recorded the spectrum with a cooled CCD coupled to a single grating spectrometer (iDus 416 and Shamrock, Andor).

We conclude that between 40-60% of the light is transmitted through the semitransparent Au substrate and between 10-40% through the Au disks, see **Figure S1**.

## 2. Raman spectra.

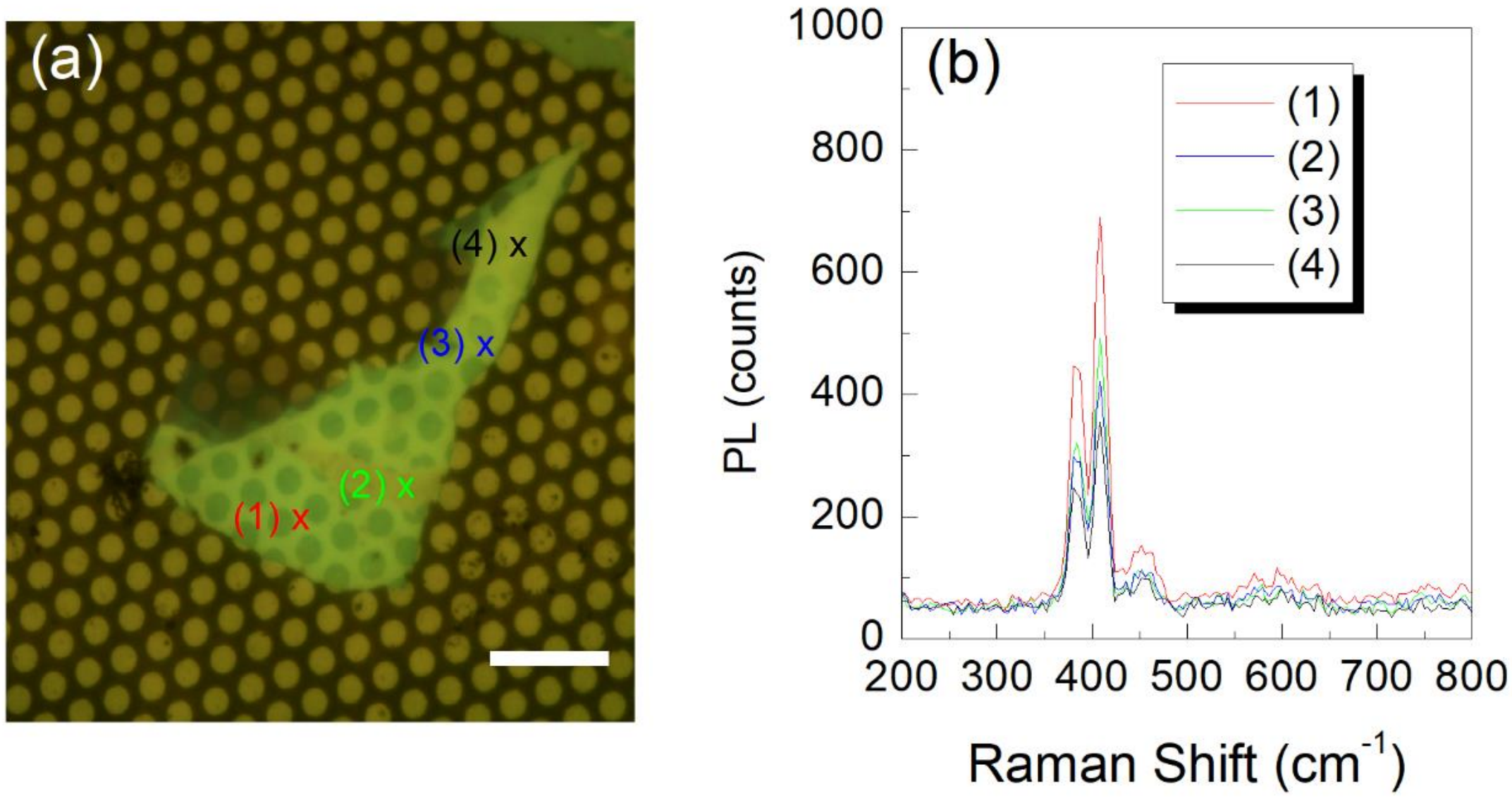

**Figure S2**. Raman spectra. (a) Optical microscopy image of the photodetectors. Scale bar is 20 µm. (b) $MoS_2$ raman spectra at the spots marked in (a).

**Figure S2b** shows Raman spectra at four different positions of the $MoS_2$ flake with a thickness ranging from 20 to 30 nm from the Figure S2a. We identify two prominent peaks: the $E^1_{2g}$ vibration mode at 383 cm$^{-1}$ and the $A_{1g}$ vibration mode at 408 cm$^{-1}$, which are in agreement with the bulk values of $MoS_2$ .[2] A 532 nm continuous wave laser (Spectra-Physics) and a cooled CCD camera (iDus 416, Andor) were used for micro-Raman measurements. The maximum power at $MoS_2$ flakes was 1 mW to prevent overheating or laser damage.

## 3. Photoluminescence

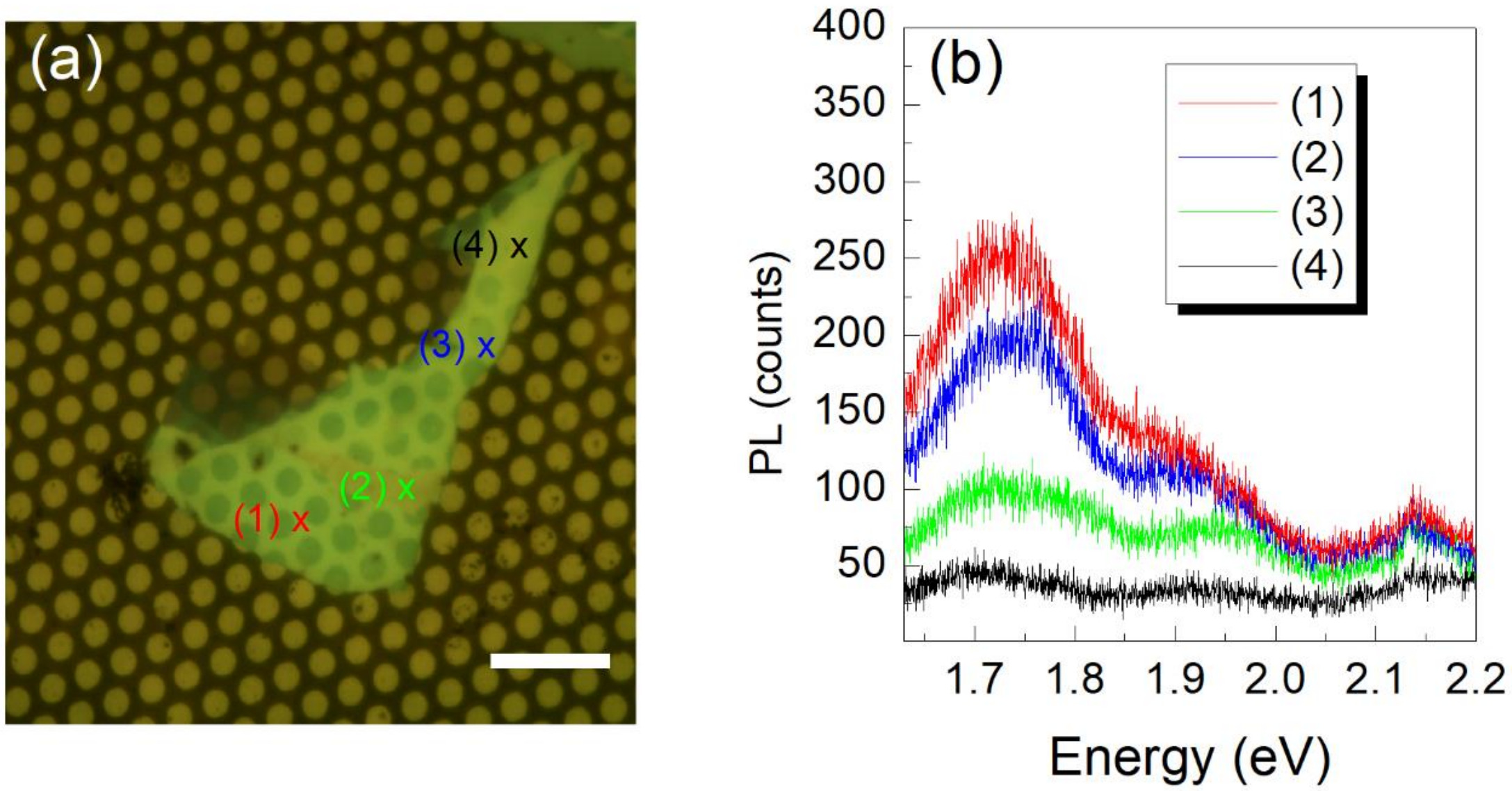

**Figure S3**. Photoluminescence. (a) Optical microscopy image of the photodetectors. Scale bar is 20 µm. (b) $MoS_2$ photoluminescence at the spots marked in (a).

Photoluminescence in few-layer $MoS_2$ is observable but the peak intensity decreases with the thicknesses of the flake.[3] We identify a prominent photoluminescence peak above 1.7 eV related to the band gap energy of $MoS_2$, see **Figure S3**. In Figure 3d in the main text, we identify an increase of responsivity above 1.7 eV, which is in agreement with these photoluminescence peaks. A smaller photoluminescence peak related to the Au substrate is identified at 2.15 eV.

## 4. Atomic Force Microscopy profile of the vertical photodetectors

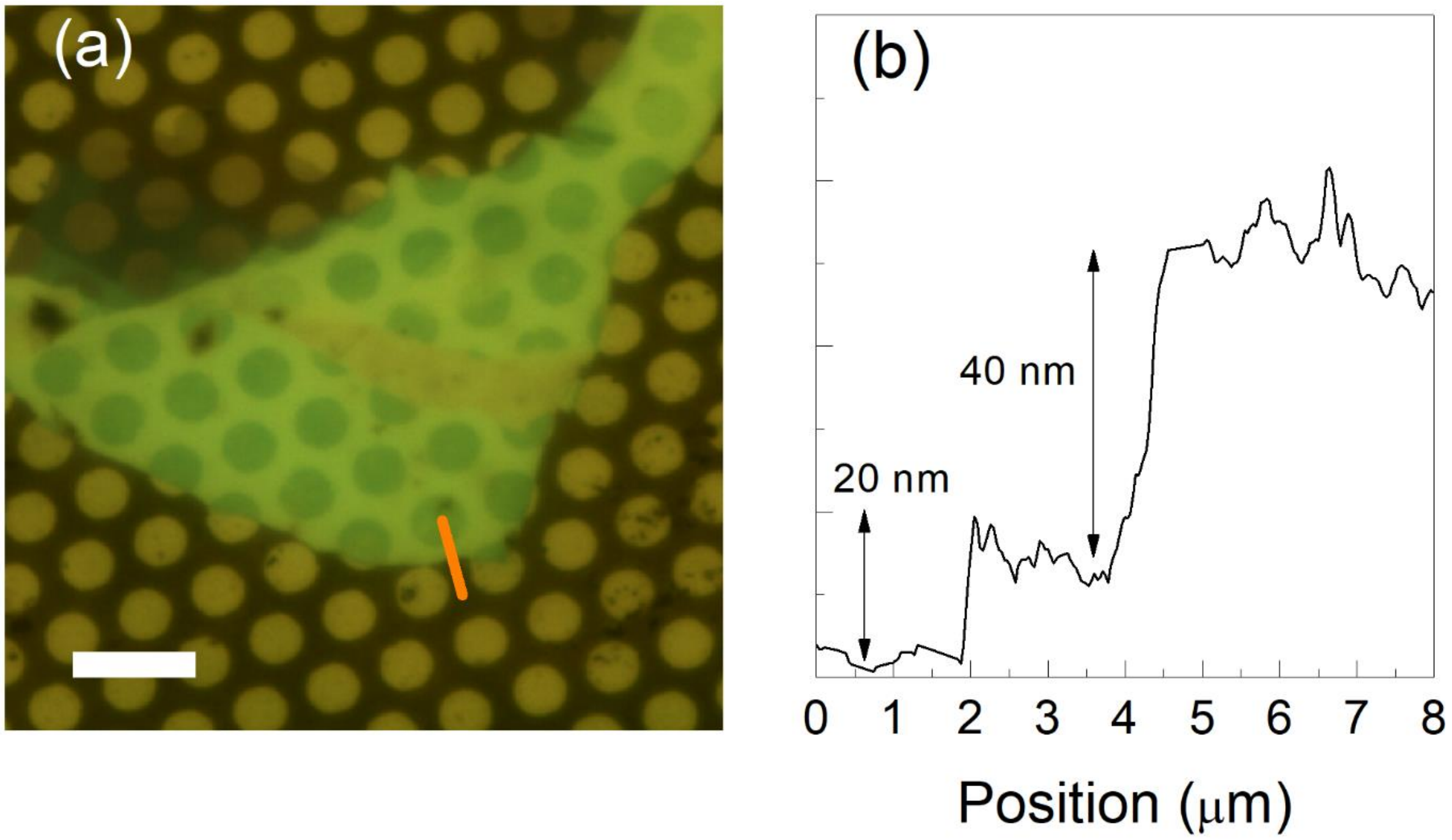

Fig. S4. Atomic Force Microscopy of the vertical $MoS_2$ photodetectors. (a) Optical microscopy image of the photodetectors. Scale bar is 10 μm. (b) Atomic Force microscopy profile across the orange line in panel (a).

We measure the thickness of the vertical $MoS_2$ photodetectors using an Atomic Force Microscopy (AFM) in contact mode to avoide possible artefacts. We measure a $MoS_2$ thickness of 20 nm and a thickness of 40 nm for the Au disk electrodes (**Figure S4**).

## 5. Comparison of $MoS_2$ Photodetectors

**Table S1:** Comparison of $MoS_2$ photodetectors

| Device configuration | Layer number | Response time (s) | Responsivity (A/W) | $V_{gate}$ (V) | Ref. |
|---|---|---|---|---|---|
| Vertical | Few-layer | $60\cdot10^{-9}$ | 0.11 | 0 | This work |
| In-plane | 1L | $2\cdot10^{-6}$ | 15 | 8 | Ref. [4] |
| Vertical | Few-layer | $7\cdot10^{-6}$ | $10\cdot10^{-3}$ | 0 | Ref. [5] |
| In-plane | Few-layer | $30\cdot10^{-6}$ | --- | 0 | Ref. [6] |
| In-plane | 3L | $40\cdot10^{-6}$ | 1.04 | --- | Ref. [7] |
| In-plane | 3L | $70\cdot10^{-6}$ | 0.57 | 0 | Ref. [8] |
| In-plane | 1L | $10\cdot10^{-3}$ | 10 | -40 | Ref. [9] |
| In-plane | 1L | $50\cdot10^{-3}$ | $7.5\cdot10^{-3}$ | 50 | Ref. [10] |
| In-plane | Few-layer | $50\cdot10^{-3}$ | $10^{5}$ | 0 | Ref. [11] |
| In-plane | Few-layer | 1 | 0.11 | -2 | Ref. [12] |
| In-plane | 2L | 2 | $10^{3}$ | 100 | Ref. [13] |
| In-plane | 1L | 3 | 2200 | 41 | Ref. [14] |
| In-plane | 1L | 4 | 880 | -70 | Ref. [15] |
| In-plane | 1L | 10 | $1.1\cdot10^{-3}$ | 0 | Ref. [16] |
| In-plane | 1L | --- | 3.5 | -40 | Ref. [17] |
| In-plane | Few-layer | --- | 343 | 8 | Ref. [18] |

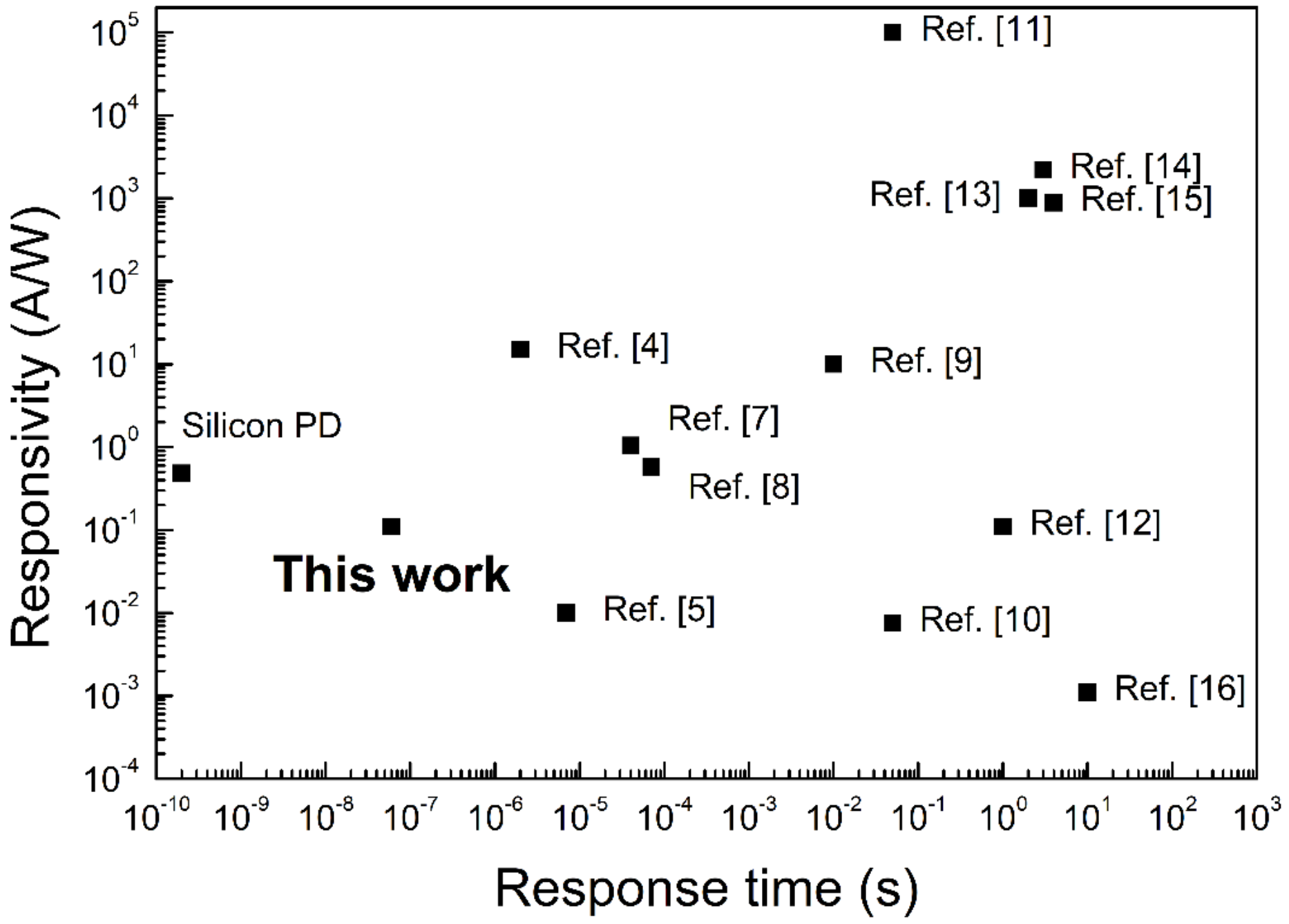

Fig. S5. Responsivity against response time for the $MoS_2$ photodevices listed in Table S1.

We present a comparison table of reported $MoS_2$ photodevices sorted by response time in **Table S1**. To give a performance overview of the $MoS_2$ photodevices, we summarized the data from Table S1 and add as a benchmark, a commercial silicon photodiode (Thorlabs, FDS02), in **Figure S5**. Our vertical photodetectors exhibit a photoresponsivity of 0.11 A/W, which is comparable to silicon technology and a fast response time for a photodetector $MoS_2$-based. This balance between responsivity and response time make our photodetectors highly efficient photocurrent devices in the visible range.

## 6. Metal-semiconductor current equation

The equation of the current $I_i$ is derived as follows: from equation (8) in page 4 of ref [19] or equation (84) in page 161 of ref [20]:

$$I = AA^* \exp(-q\phi_{SB}/k_B T)\,[\exp(qV_b/k_B T) - 1]$$

where $A$ is the area of the device, $A^*$ is the Richardson constant, $q$ is the electron charge, $\phi_{SB}$ is the Schottky barrier, $k_B$ is the Boltzmann constant, $T$ is temperature and $V_b$ the bias voltage applied.

Considering that the barrier height depends on the bias voltage applied, we include the effect of the image force as an additional barrier as Rhoderick *et al.*: $\phi_{SB} = \phi_{SB,0} + \beta V$, and the ideality factor is $1/\eta = 1 - \beta$. We can now rewrite the equation of the current:

$$I = AA^* \exp\left(-q\phi_{SB,0}/k_B T\right) \exp(-q\beta V_b/k_B T)\,[\exp(qV_b/k_B T) - 1] =$$

$$I_0 \exp(-q\beta V_b/k_B T) \exp(qV_b/k_B T)\,[1 - \exp(-qV_b/k_B T)] =$$

$$I_0 \exp(q(1-\beta)V_b/k_B T)\,[1 - \exp(-qV_b/k_B T)] =$$

$$I_0 \exp(qV_b/\eta k_B T)\,[1 - \exp(-qV_b/k_B T)]$$

where $I_0 = AA^* \exp\left(-q\phi_{SB,0}/k_B T\right)$. This is the expression used in the *I-V* characteristics fit.